\documentclass[twocolumn]{emulateapj}

\begin{document}

\title{Sublimation-Driven Activity in Main-Belt Comet 313P/Gibbs}

\author{
Henry H.\ Hsieh\altaffilmark{1},
Olivier Hainaut\altaffilmark{2},~
Bojan Novakovi\'c\altaffilmark{3},~
Bryce Bolin\altaffilmark{4},~
Larry Denneau\altaffilmark{5},~
Alan Fitzsimmons\altaffilmark{6},~
Nader Haghighipour\altaffilmark{5},~
Jan Kleyna\altaffilmark{5},~
Rosita Kokotanekova\altaffilmark{7,8},~
Pedro Lacerda\altaffilmark{8},~
Karen J.\ Meech\altaffilmark{5},~
Marco Micheli\altaffilmark{9},~
Nick Moskovitz\altaffilmark{10},~
Eva Schunova\altaffilmark{5},~
Colin Snodgrass\altaffilmark{7},~
Richard J.\ Wainscoat\altaffilmark{5},~
Lawrence Wasserman\altaffilmark{10},~
Adam Waszczak\altaffilmark{11}~
}
\altaffiltext{1}{Institute of Astronomy and Astrophysics, Academia Sinica, P.O.\ Box 23-141, Taipei 10617, Taiwan}
\altaffiltext{2}{European Southern Observatory, Karl-Schwarzschild-Stra\ss e 2, D-85748 Garching bei MŸ{\"u}nchen, Germany}
\altaffiltext{3}{Department of Astronomy, Faculty of Mathematics, University of Belgrade, Studentski trg 16, 11000 Belgrade, Serbia}
\altaffiltext{4}{Observatoire de la C{\^o}te d'Azur, Boulevard de l'Observatoire, B.P.\ 4229, 06304 Nice Cedex 4, France}
\altaffiltext{5}{Institute for Astronomy, University of Hawaii, 2680 Woodlawn Drive, Honolulu, HI 96822, USA}
\altaffiltext{6}{Astrophysics Research Centre, Queens University Belfast, Belfast BT7 1NN, United Kingdom}
\altaffiltext{7}{Planetary and Space Sciences, Department of Physical Sciences, The Open University, Milton Keynes, MK7 6AA, United Kingdom}
\altaffiltext{8}{Max Planck Institute for Solar System Research, Justus-von-Liebig-Weg 3, 37077 G{\"o}ttingen, Germany}
\altaffiltext{9}{ESA SSA NEO Coordination Centre, Frascati, RM, Italy}
\altaffiltext{10}{Lowell Observatory, 1400 W.\ Mars Hill Road, Flagstaff, AZ 86001, USA}
\altaffiltext{11}{Division of Geological and Planetary Sciences, California Institute of Technology, Pasadena, CA 91125}
\email{hhsieh@asiaa.sinica.edu.tw}

\slugcomment{ApJ Letters - Updated 2014 Nov 18}

\begin{abstract}
We present an observational and dynamical study of newly discovered main-belt comet 313P/Gibbs.  We find that the object is clearly active both in observations obtained in 2014 and in precovery observations obtained in 2003 by the Sloan Digital Sky Survey, strongly suggesting that its activity is sublimation-driven.  This conclusion is supported by a photometric analysis showing an increase in the total brightness of the comet over the 2014 observing period, and dust modeling results showing that the dust emission persists over at least three months during both active periods, where we find start dates for emission no later than 2003 July 24$\pm$10 for the 2003 active period and 2014 July 28$\pm$10 for the 2014 active period.  From serendipitous observations by the Subaru Telescope in 2004 when the object was apparently inactive, we estimate that the nucleus has an absolute $R$-band magnitude of $H_R=17.1\pm0.3$, corresponding to an effective nucleus radius of $r_e\sim1.00\pm0.15$~km.  The object's faintness at that time means we cannot rule out the presence of activity, and so this computed radius should be considered an upper limit.  We find that 313P's orbit is intrinsically chaotic, having a Lyapunov time of $T_l=12\,000$~yr and being located near two 3-body mean-motion resonances with Jupiter and Saturn, 11J-1S-5A and 10J+12S-7A, yet appears stable over $>$50~Myr in an apparent example of stable chaos.  We furthermore find that 313P is the second main-belt comet, after P/2012 T1 (PANSTARRS), to belong to the $\sim$155~Myr~old Lixiaohua asteroid family.
\end{abstract}


\newpage

\section{INTRODUCTION}

Comet 313P/Gibbs was discovered as P/2014 S4 (Gibbs) on 2014 September 24 by A.~R.\ Gibbs using the Catalina Sky Survey's 0.68-m Schmidt telescope.  Follow-up images from the Mt.\ Lemmon 1.5~m telescope revealed a 20$''$ tail and a slightly elliptical coma with a full-width at half-maximum (FWHM) approximately twice that of nearby field stars \citep{gib14}.  Analysis of observations by the Hubble Space Telescope (HST), Keck I, and the Danish 1.5~m telescope at La Silla indicate continuous mass loss between 2014 October 2 and 2014 November 6, inconsistent with an impact event, though no spectroscopic evidence of gas emission was detected \citep{jew15}.  As of 2015 January  15, 313P has an osculating semimajor axis, eccentricity, and inclination of $a=3.156$~AU, $e=0.242$, and $i=10.97^{\circ}$, respectively, an orbital period of 5.61~yr, and a Tisserand parameter with respect to Jupiter of $T_J=3.132$.  These orbital elements place the object unequivocally within the main asteroid belt, making it one of the growing number of active asteroids \citep[cf.][]{jew12} in the main belt to be discovered in recent years.

Active asteroids include main-belt comets \citep[MBCs; cf.][]{hsi06}, which exhibit cometary activity due to ice sublimation, and disrupted asteroids \citep[cf.][]{hsi12a}, where dust emission is caused by effects such as impacts or rotational disruptions.  Distinguishing between different sources of activity is crucial for developing our understanding of the global properties (e.g., abundances, spatial distributions, and physical characteristics) of objects exhibiting each type of activity \citep[e.g.,][]{hsi15}.

Unambiguously determining the source of comet-like activity, especially for a recently discovered object, is unfortunately not straightforward.  Despite detections of water ice frost on asteroid (24) Themis \citep{riv10,cam10} and water vapor outgassing from (1) Ceres \citep{kup14}, attempts to directly detect gas emission from active asteroids via spectroscopy \citep[e.g.,][]{jew09,lic11,hsi12a,hsi12b,hsi12c,hsi13,dev12,oro13} have all been unsuccessful.  We must use indirect methods instead, like dust modeling or photometric monitoring \citep[cf.][]{hsi12a}, to determine whether dust emission was long-lived \citep[implying sublimation-driven emission; e.g.,][]{hsi12b} or impulsive \citep[implying impact-driven ejection; e.g.,][]{ste12}.  As such, unambiguous confirmation of sublimation is difficult to achieve, a situation further complicated in cases where multiple mechanisms could be at work \citep[e.g.,][]{jew14a}.  Recurrent activity near perihelion, such as that observed for MBCs 133P/Elst-Pizarro \citep{hsi04,hsi10,jew14b} and 238P/Read \citep{hsi11b}, is currently considered the most reliable indicator of sublimation-driven activity. This behavior is naturally explained by thermally-modulated sublimation and extremely difficult to explain as a consequence of any physical disruption \citep{hsi12a,jew12}.

\begin{figure*}
\plotone{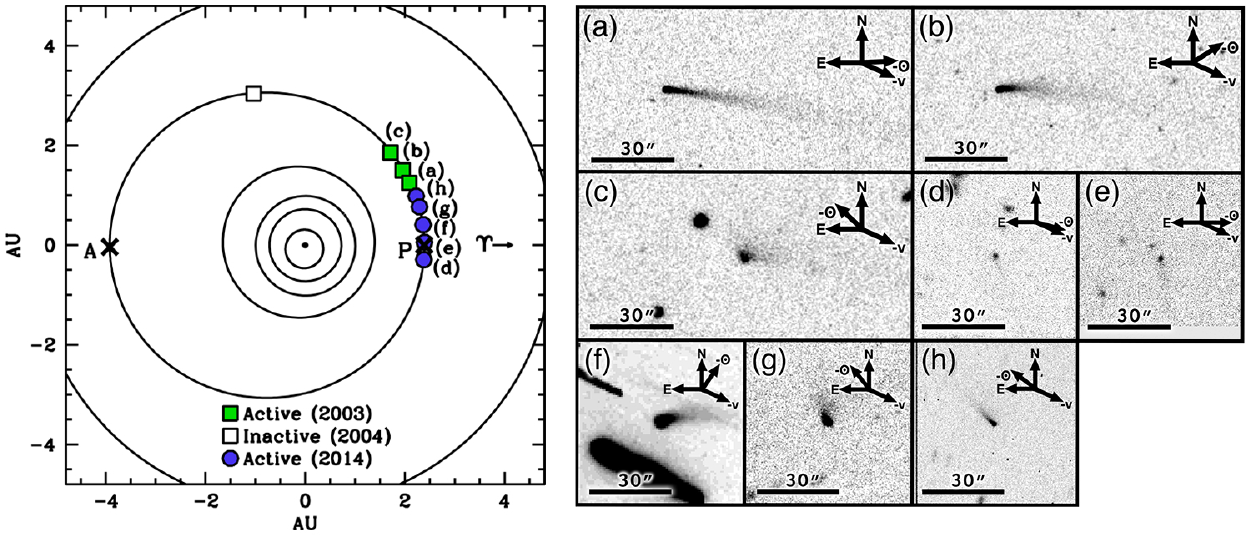}
\caption{\small 
Orbital position plot (left) and composite images (right) constructed from observations detailed in Table~\ref{obslog}, with the Sun (black dot) at the center of the orbital position plot, and the orbits of Mercury, Venus, Earth, Mars, 313P, and Jupiter shown as black lines.  Perihelion (P) and aphelion (A) are marked with crosses.   Colored points (left panel) and images (right panel) correspond to observations from (a) 2003 September 30, (b) 2003 October 23, (c) 2003 November 28, (d) 2014 August 6, (e) 2014 September 3, (f) 2014 October 1, (g) 2014 October 30, and (h) 2014 November 18.  
An open square symbol (left panel) marks the position of observations from 2004 September 16 (image not shown).
The object is at the center of each image panel, and North (N), East (E), the antisolar vector ($-\odot$), and the negative heliocentric velocity vector ($-v$) are marked with arrows.
  }
\label{figure:fig_images_orbplot}
\end{figure*}

\section{OBSERVATIONS\label{observations}}

Following 313P's discovery, we identified pre-discovery images from August 2014 obtained by the 1.8~m Pan-STARRS1 (PS1) survey telescope using PS1 $i_{\rm P1}$- and $w_{\rm P1}$-band filters \citep{ton12}.
We also obtained follow-up imaging in October and November 2014 using
the Large Format Camera \citep{sim00} on Palomar Observatory's 5.1~m Hale telescope,
the Large Monolithic Imager \citep{bid14} on Lowell Observatory's 4.3-m Discovery Channel Telescope,
a $4096\times4096$ Fairchild CCD on the 2.0~m Faulkes Telescope North on Haleakala,
a $2048\times2048$ Textronix CCD on the University of Hawaii (UH) 2.2~m telescope on Mauna Kea,
and the European Southern Observatory (ESO) Faint Object Spectrograph and Camera \citep{buz84} on the 3.58~m New Technology Telescope (Program 194.C-0207) operated by ESO at La Silla.
SDSS $r'$-band or Kron-Cousins $R$-band filters, as specified in Table~\ref{obslog}, were used for all follow-up observations.
We used Image Reduction and Analysis Facility software \citep{tod86} to perform standard bias subtraction and flat-field reduction (using dithered twilight sky images) for all data, except those from PS1, which were reduced using the system's Image Processing Pipeline \citep{mag06}.

\setlength{\tabcolsep}{3.0pt}
\begin{deluxetable*}{lcrrcrrrrcccc}
\scriptsize
\tablewidth{0pt}
\tablecaption{Observations\label{obslog}}
\tablecolumns{14}
\tablehead{
\colhead{UT Date}
 & \colhead{Tel.\tablenotemark{a}}
 & \colhead{N\tablenotemark{b}}
 & \colhead{t\tablenotemark{c}}
 & \colhead{Filter}
 & \colhead{$R$\tablenotemark{d}}
 & \colhead{$\Delta$\tablenotemark{e}}
 & \colhead{$\alpha$\tablenotemark{f}}
 & \colhead{$\nu$\tablenotemark{g}}
 & \colhead{$m_R(R,\Delta,\alpha)$\tablenotemark{h}}
 & \colhead{$m_R(1,1,0)$\tablenotemark{i}}
 & \colhead{$m_{R,{\rm tot}}(1,1,0)$\tablenotemark{j}}
 & \colhead{Active?\tablenotemark{k}}
}
\startdata
2003 June 20      & \multicolumn{4}{l}{\it Perihelion.......................} & 2.367 & 2.604 & 23.0 &   0.0 & --- & --- & --- & --- \\ 
2003 September 30 & SDSS   &  4 &  216 & $g'r'i'z'$    & 2.434 & 1.598 & 16.1 &  30.5 & 18.4$\pm$0.1 & 14.6$\pm$0.1 & 13.6$\pm$0.1 & yes \\ 
2003 October 23   & SDSS   &  4 &  216 & $g'r'i'z'$    & 2.465 & 1.512 &  8.5 &  37.0 & 18.2$\pm$0.1 & 14.8$\pm$0.1 & 13.9$\pm$0.1 & yes \\ 
2003 October 24   & SDSS   &  4 &  216 & $g'r'i'z'$    & 2.467 & 1.511 &  8.3 &  37.2 & 18.3$\pm$0.1 & 14.9$\pm$0.1 & 14.1$\pm$0.1 & yes \\ 
2003 November 28  & SDSS   &  4 &  216 & $g'r'i'z'$    & 2.525 & 1.636 & 12.1 &  46.7 & 18.7$\pm$0.1 & 14.9$\pm$0.1 & 14.2$\pm$0.1 & yes \\ 
2004 September 16 & Subaru &  2 &  120 & $W$-$J$-$VR$  & 3.208 & 3.742 & 14.1 & 108.7 & 23.3$\pm$0.2 & 17.1$\pm$0.2 & ---          & no  \\ 
2009 January 18   & \multicolumn{4}{l}{\it Perihelion.......................} & 2.385 & 2.977 & 17.0 &   0.0 & --- & --- & --- & --- \\ 
2014 August 29    & \multicolumn{4}{l}{\it Perihelion.......................} & 2.392 & 1.555 & 16.9 &   0.0 & --- & --- & --- & --- \\ 
2014 August 06    & PS1    &  2 &   90 & $w_{\rm P1}$ & 2.395 & 1.750 & 22.1 & 353.5 & 20.8$\pm$0.1 & 16.6$\pm$0.1 & 16.6$\pm$0.1 & yes \\ 
2014 August 18    & PS1    &  3 &  135 & $i_{\rm P1}$ & 2.392 & 1.637 & 19.6 & 357.0 & 20.6$\pm$0.1 & 16.6$\pm$0.1 & 16.6$\pm$0.1 & yes \\ 
2014 September 03 & PS1    &  2 &   90 & $w_{\rm P1}$ & 2.392 & 1.518 & 15.2 &   1.8 & 20.3$\pm$0.1 & 16.7$\pm$0.1 & 16.7$\pm$0.1 & yes \\ 
2014 October 01   & P200   & 38 & 1140 & $r'$         & 2.399 & 1.429 &  7.7 &  10.0 & 19.3$\pm$0.1 & 16.1$\pm$0.1 & 15.8$\pm$0.1 & yes \\ 
2014 October 13   & FTN    &  2 &  120 & $r'$         & 2.404 & 1.444 &  8.4 &  13.2 & 19.4$\pm$0.1 & 16.1$\pm$0.1 & 16.0$\pm$0.1 & yes \\ 
2014 October 17   & DCT    & 29 & 2610 & $r'$         & 2.407 & 1.461 &  9.6 &  14.7 & 19.3$\pm$0.1 & 15.9$\pm$0.1 & 15.6$\pm$0.1 & yes \\ 
2014 October 22   & DCT    &  2 &  240 & $R$          & 2.410 & 1.484 & 11.0 &  16.2 & 19.6$\pm$0.1 & 16.1$\pm$0.1 & 15.8$\pm$0.1 & yes \\
2014 October 30   & UH2.2  &  5 & 1500 & $R$          & 2.416 & 1.530 & 13.4 &  18.5 & 19.6$\pm$0.1 & 16.0$\pm$0.1 & 15.8$\pm$0.1 & yes \\ 
2014 November 17  & NTT    &  4 &  480 & $r'$         & 2.432 & 1.683 & 18.4 &  23.7 & 20.2$\pm$0.1 & 16.2$\pm$0.1 & 16.0$\pm$0.1 & yes \\ 
2014 November 18  & NTT    &  3 &  900 & $r'$         & 2.433 & 1.694 & 18.6 &  24.0 & 20.6$\pm$0.1 & 16.6$\pm$0.1 & 15.7$\pm$0.1 & yes \\ 
2014 December 16  & UH2.2  &    &      & $R$          & 2.464 & 2.023 & 22.7 &  31.9 &              &              &              & yes \\ 
2020 April 15     & \multicolumn{4}{l}{\it Perihelion.......................} & 2.419 & 3.344 &  7.8 &   0.0 & --- & --- & --- & --- 
\enddata
\tablenotetext{a}{Telescope (SDSS: Sloan Digital Sky Survey; PS1: Pan-STARRS1; P200: Palomar Hale Telescope; FTN: Faulkes Telescope North; DCT: Discovery Channel Telescope; UH2.2: UH 2.2~m telescope; NTT: New Technology Telescope).}
\tablenotetext{b}{Number of exposures.}
\tablenotetext{c}{Total integration time, in s.}
\tablenotetext{d}{Heliocentric distance, in AU.}
\tablenotetext{e}{Geocentric distance, in AU.}
\tablenotetext{f}{Solar phase angle (Sun-object-Earth), in degrees.}
\tablenotetext{g}{True anomaly, in degrees.}
\tablenotetext{h}{Mean apparent $R$-band magnitude, assuming solar colors.}
\tablenotetext{i}{Absolute $R$-band magnitude (at $R=\Delta=1$~AU and $\alpha=0\degr$) as measured with a $5\farcs0$ aperture, assuming IAU $H,G$ phase-darkening where $G=0.15$.}
\tablenotetext{j}{Absolute $R$-band magnitude as measured with rectangular aperture enclosing entire visible extent of comet.}
\tablenotetext{k}{Is visible activity present?}
\end{deluxetable*}

Using the Solar System Object Image Search tool \citep[SSOIS;][]{gwy12}, provided by the Canadian Astronomical Data Centre, we identified precovery observations obtained in 2003 by the 2.5~m Sloan Digital Sky Survey (SDSS) telescope \citep{yor00,fuk96,gun98,gun06,aih11}, and in 2004 by Suprime-Cam \citep{miy02} on the 8.2~m Subaru Telescope.  The SDSS observations, obtained in $g'$-, $r'$-, $i'$-, and $z'$-band filters, show a clearly cometary object with a dust tail extending as much as $\sim$1~arcmin in the antisolar direction.  The object is extremely faint in the Subaru images, obtained using a wide-band ``W-J-VR'' filter encompassing both the Johnson V and Cousins R passbands, exhibiting no indications of activity.

Absolute calibration of all data was accomplished using magnitudes of field stars with approximately solar colors from SDSS \citep{aih11} or PS1 \citep{sch12,ton12,mag13}.  Conversion of all photometry to $R$-band was accomplished using transformations derived by \citet{ton12} and by R.\ Lupton ({\tt http://www.sdss.org/}).  Comet photometry was performed using circular apertures with varying radii depending on the nightly seeing, where background statistics were measured in nearby regions of blank sky, avoiding contamination from the comet.

\section{RESULTS\label{section:results}}

\subsection{Photometric Analysis\label{section:phot_results}}

\begin{figure*}
\plotone{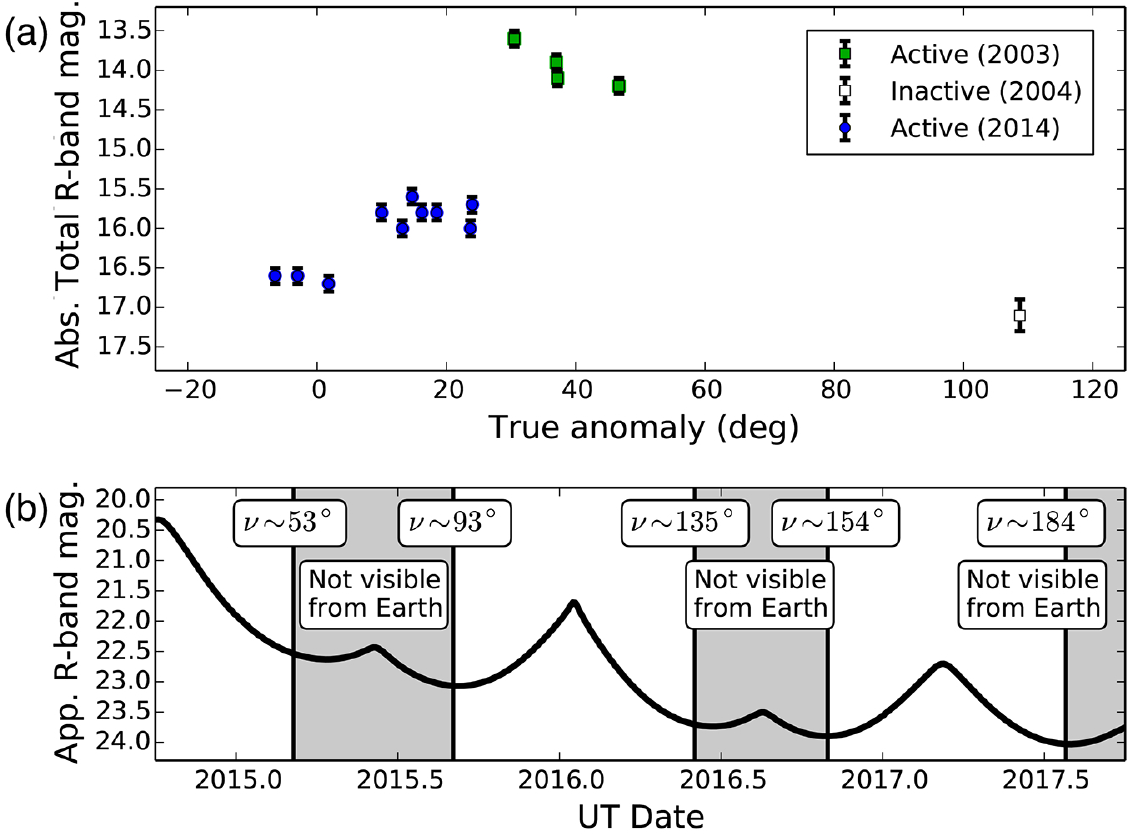}
\caption{\small (a) Plot of absolute total R-band magnitudes measured for 313P (including the total observed flux from the nucleus, coma, and dust tail) from 2003 through 2014. (b) Predicted apparent R-band magnitude of the nucleus of 313P through the end of the 2016-2017 observing window.  Shaded regions mark time periods when the object is not observable from Earth (i.e., solar elongation $<$50$^{\circ}$).
  }
\label{figure:fig_photevolution_visibility}
\end{figure*}

Apparent $R$-band magnitudes as measured in 5$''$ circular apertures, and equivalent absolute magnitudes (assuming a standard $H,G$ phase function where we assume $G=0.15$), as well as absolute magnitudes corresponding to larger rectangular apertures encompassing the entire comet, for all observations are listed in Table~\ref{obslog}.  Both the central and total brightnesses of the comet decline over the 2003 observing period (which spans a true anomaly range of $30.5^{\circ}<\nu<46.7^{\circ}$), declining by $\sim$40\% in 59 days. Meanwhile, the comet brightens over the 2014 observing period ($-6.5^{\circ}<\nu<24.0^{\circ}$), increasing in total brightness by $\sim$150\% in 104 days (Figure~\ref{figure:fig_photevolution_visibility}a).  The object's increasing brightness in 2014 is suggestive of ongoing, sublimation-driven dust production, but is not unambiguous proof of sublimation \citep[cf.][]{hsi12a}. Multi-filter SDSS observations from 2003 show that the comet has approximately solar colors ($g'-r'=0.54\pm0.05$, $r'-i'=0.10\pm0.05$, and $i'-z'=0.05\pm0.05$; equivalent to $B-V=0.71\pm0.05$, $V-R=0.41\pm0.05$, and $R-I=0.33\pm0.05$) with no appreciable color differences between the coma and the tail, suggesting that the measured colors in both regions are dominated by dust.  Strong coma during both active periods prevent any meaningful determination of a rotational period or amplitude.

We measure an apparent $R$-band magnitude of $m_R=23.3\pm0.2$ for 313P in 2004, when it was apparently inactive, while it was at $\nu=108.7^{\circ}$ and at a heliocentric distance of $R=3.21$~AU.  This corresponds to an absolute magnitude of $H_R=17.1\pm0.2$, from which we compute an effective nucleus radius of $r_e\sim1.0\pm0.1$~km \citep[assuming a $R$-band albedo of $p_R=0.05$; cf.][]{hsi09}, making it a mid-sized MBC \citep[cf.][]{hsi15}.  While no coma or tail was visible in this data, the object's faintness at the time means that we cannot rule out the presence of activity, particularly since activity was detected for MBCs 133P and P/2010 R2 (La Sagra) as late as $\nu=109^{\circ}$ and $\nu=117^{\circ}$, respectively, when both were at $R\sim3.25$~AU \citep{kal11,hsi14}.  In fact, HST observations indicate 313P's nucleus may be as small as $r_e\sim0.5$~km \citep{jew15}.  As such, the nucleus size calculated here should be regarded as an upper limit.

Using this nucleus size, we can estimate total dust masses at different times.  Following \citet{hsi14}, we find a peak total dust mass of $M_d\approx(2.5\pm1.0)\times10^8$~kg and dust-to-nucleus mass ratio of $M_d/M_N\approx(4.5\pm1.5)\times10^{-5}$ in 2003 (at $\nu=30.5^{\circ}$) and $M_d\approx(3\pm1)\times10^7$~kg and $M_d/M_N\approx(5.5\pm2.0)\times10^{-6}$ in 2014 (at $\nu=24^{\circ}$), or a decline in peak total dust mass of $\sim1$ order of magnitude between 2003 and 2014.  Additional photometric measurements through the end of the 2014-2015 observing window will permit a more direct comparison of 2003 and 2014-2015 activity levels over the same orbital arcs, while more accurate measurements of 313P's nucleus size will improve the accuracy of the dust mass estimates reported here. However, these current results suggest that there is a significant decline in activity strength after just two orbit passages, perhaps due to a combination of mantling of the active area responsible for the observed activity \citep[cf.][]{hsi15} and depletion of exposed volatile material.  Alternatively, other effects (e.g., outbursts or pole orientation evolution) could be responsible, and as such, monitoring of future active episodes is highly encouraged to constrain possible causes for the observed decline in activity strength.  Generally speaking, however, 313P's activity levels in both 2003 and 2014 are comparable to those of other MBCs both in terms of peak total dust masses and peak dust-to-nucleus mass ratios \citep[cf.][]{hsi14}.

Our nucleus size estimate also permits us to predict the comet's brightness in the absence of activity (or at least at the level of activity exhibited in 2004) in the future (Figure~\ref{figure:fig_photevolution_visibility}b).  Observed magnitudes fainter than predicted will indicate that the nucleus smaller than estimated here (implying that residual activity was present in 2004), while magnitudes brighter than predicted will indicate the presence of activity.

\subsection{Morphological Analysis\label{section:dustmodel_results}}

Highly variable morphology was observed for 313P in both 2003 and 2014.  The projected orientation of the dust tail in the sky rotates counterclockwise during both active periods, each time lagging slightly behind the similarly rotating antisolar vector (as projected on the sky), exhibiting strong similarities to the morphological evolution of P/La Sagra in 2010 \citep{hsi12c}.  For P/La Sagra, two dust tails (one oriented in the antisolar direction, and another aligned with the object's orbit plane) were observed late in its observed active period \citep{hsi12c}.  This distinctive morphology was also observed for 288P \citep{hsi12b} and was interpreted to indicate long-lived dust emission, since the antisolar dust tail indicated the presence of small, fast-moving dust particles that must have been recently emitted, while the orbit-plane-aligned dust tail corresponded to large, slow-moving particles that must have been ejected much longer ago.  Only one tail has been observed for 313P to date, but if the observed activity is in fact sublimation-driven, further observations could eventually reveal the development of a second dust tail.

\begin{figure*}
\plotone{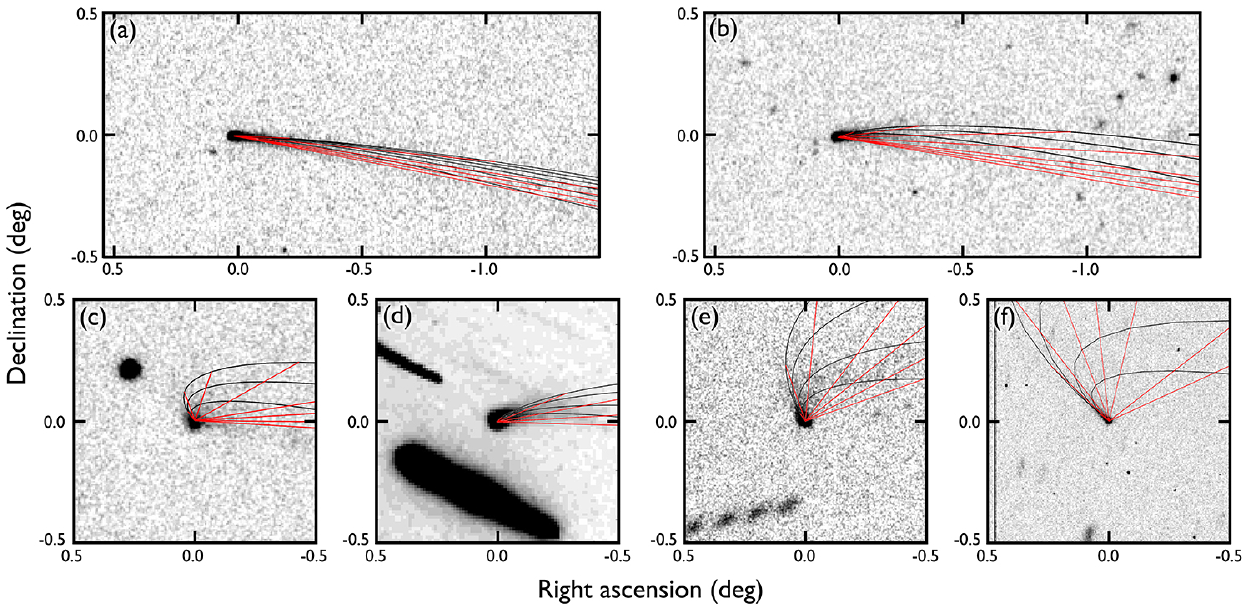}
\caption{\small 
Syndynes (black lines) and synchrones (red lines) overplotted on data from (a) 2003 September 30, (b) 2003 October 23, (c) 2003 November 28, (d) 2014 October 1, (e) 2014 October 30, and (f) 2014 November 18, where $\beta=0.03$ syndynes and $t_{\rm ej}=-10$~day synchrones are the upper-most or left-most black and red lines, respectively, in each panel.
  }
\label{figure:fig_fpmodels_images}
\end{figure*}

To quantitatively analyze 313P's dust morphology, we perform a \citet{fin68} numerical dust modeling analysis.  In this formalism, dust grains are released with zero velocity from the nucleus some time in the past, and their position is computed at the time of observations, accounting for solar gravity and solar radiation pressure. A range of ejection times and particle sizes are considered, where particle sizes are parametrized using the ratio of solar radiation pressure to solar gravity, $\beta$, where $1/\beta$ gives approximate particle radii, $a$, in microns. The locus of the present positions of all particles with a common ejection time is called a synchrone, while the locus of the present positions of all particles with a common $\beta$ is a syndyne.

On each image, no gap is observed between the dust tail and the nucleus, implying either that (1) a continuum of dust grain sizes up to very large grains is present, where in practice, the seeing disk prevents us from setting an upper limit to the grain size, or (2) there has been continued emission of dust from some time in the past until the present, where the seeing disk does not allow us determine whether the emission has very recently stopped.  The envelope of the coma is effectively enclosed by a syndyne representing $\beta_{\rm max}$ (the minimum particle size) and a synchrone representing $t_{\rm ej,min}$ (the earliest ejection time), such that all the visible dust has $\beta<\beta_{\rm max}$ and was ejected at $t_{\rm ej}>t_{\rm ej,min}$.  The values of $\beta_{\rm max}$ and $t_{\rm ej,min}$ themselves represent lower and upper limits, respectively, since observations with higher $S/N$ could reveal fainter surface brightness features, thus extending the envelope of the observed coma and tail.

In Figure~\ref{figure:fig_fpmodels_images}, we overplot syndynes (black lines) representing particles sizes ranging from $\beta=0.005$ to $\beta=0.03$ ($a\sim30-200$~$\mu$m) and synchrones (red lines) representing ejection times from $t_{\rm ej}=-10$ days to $t_{\rm ej}=-130$ days (with respect to the time of observations) on images from Figure~\ref{figure:fig_images_orbplot}.
We find good agreement among our models of individual observations indicating $\beta_{\rm max}\geq0.03$ (equivalent to particles tens of microns in radius, and larger), and emission start times of no later than 2003 July 24$\pm$10 ($\nu\lesssim10^{\circ}\pm3^{\circ}$) for the 2003 activity, and no later than 2014 July 28$\pm$10 ($\nu\lesssim351^{\circ}\pm3^{\circ}$) for the 2014 activity, i.e., with activity persisting for $>$3 months each time.  We find no evidence of significant changes (i.e., outbursts or quiescent periods) in the intensity of the activity or in the dust particle size distribution over the emission period.

\subsection{Dynamical Analysis\label{dyn_results}}

\begin{figure*}
\plotone{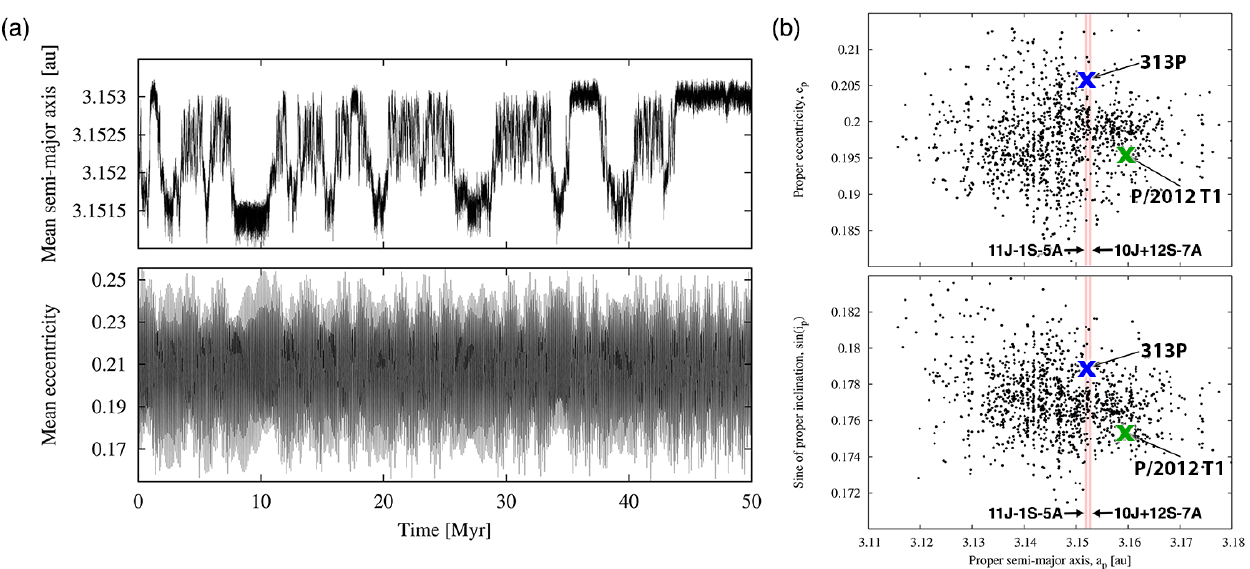}
\caption{\small (a) Evolution of mean semimajor axis (top) and mean eccentricity (bottom) of 313P over 50 Myr from numerical integrations. (b) Plots of proper semimajor axis versus proper eccentricity (top) and sine of proper inclination (bottom) for Lixiaohua family members (black dots) and MBCs 313P (blue X symbol) and P/2012 T1 (green X symbol).  Vertical red shaded regions mark the 11J-1S-5A and 10J+12S-7A MMRs.
  }
\label{figure:fig_dynamics}
\end{figure*}

To ascertain some of 313P's basic dynamical properties, we first numerically propagate its orbit for 10 Myr using the Orbit9 integrator ({\tt http://hamilton.dm.unipi.it/astdys/}), accounting for the four outer planets (Jupiter to Neptune).  We calculate proper orbital elements of $a_p = 3.15204$~AU, $e_p = 0.20560$, and $i_p = 10.30^{\circ}$, and a Lyapunov time of $T_l = 12\,000$ yr. This short $T_l$ indicates that 313P's orbit is intrinsically chaotic, perhaps due to the proximity of two 3-body mean motion resonances (MMRs) with Jupiter and Saturn, 11J-1S-5A and 10J+12S-7A. These two resonances are relatively weak, but have overlapping zones of influence for $e_p > 0.15$, forming a highly chaotic zone around 313P.

Next, we extend our integrations to 50 Myr, aiming to better understand 313P's long-term dynamical behavior. We find that the object's orbital elements remain nearly constant over even this longer integration period (Figure~\ref{figure:fig_dynamics}a). Thus, despite residing in a chaotic region of the asteroid belt, the orbit of 313P appears relatively stable. Such behavior has also been observed for other asteroids in high-order MMRs \citep{tsi02}, illustrating a phenomenon known as stable chaos \citep{mil92}.

We also search for an associated dynamical family to check whether 313P may have originated in a recent fragmentation event like 133P and 288P \citep{nes08,nov12}. Employing the hierarchical clustering method \citep{zap90}, we find that 313P is associated with the $\sim$155 Myr old Lixiaohua asteroid family \citep[Figure~\ref{figure:fig_dynamics}b;][]{nov10}, joining the family at the low cut-off velocity of $d\sim20$~m~s$^{-1}$, making it the second MBC in this family after P/2012 T1 \citep{hsi13}. All Lixiaohua members, including 313P, have similarly chaotic orbits \citep{nov10}, and as such, 313P's Lixiaohua family association should be considered as reliable as those of other members, despite the chaoticity of its orbit inferred here.  Unfortunately, as in the case of P/2012 T1, any search for associations with younger families will be complicated by the high number density of asteroids in this region of orbital element space, as well as the region's intrinsic chaoticity.

\section{DISCUSSION\label{discussion}}

After 133P \citep{hsi04,hsi10,jew14b} and 238P \citep{hsi11b}, 313P is now the third MBC observed to exhibit recurrent activity.  We conclude from this evidence, along with photometric and dust modeling results presented here, that 313P's activity is likely sublimation-driven, perhaps due to the collisional excavation of subsurface ice \citep[e.g.,]{hsi04,cap12}, consistent with the conclusions of \citet{jew15}.

Preliminary results from numerical dynamical modeling indicate that some main-belt objects with moderately large eccentricities and inclinations similar to those of 313P, particularly those near mean-motion resonances, may be temporarily-captured JFC-like interlopers, even if they have $T_J\gg3$ \citep{hsi14b}.  Therefore, while our integrations of 313P itself indicate that it is moderately stable, and thus possibly native to the main belt, we caution that its dynamical origin cannot be considered absolutely certain.

Continued observations of 313P are encouraged until March 2015 (the end of the 2014-2015 observing window; Figure~\ref{figure:fig_photevolution_visibility}) to continue monitoring its photometric and morphological evolution.  From September 2015 to May 2016, 313P will have a true anomaly range of $93^{\circ}<\nu<135^{\circ}$, and may provide opportunities to characterize its inactive nucleus.  However, given aforementioned examples of MBCs remaining active over similar true anomaly ranges (Section~\ref{section:phot_results}), observations from November 2016 to July 2017 ($154^{\circ}<\nu<184^{\circ}$), may prove more reliable for nucleus characterization purposes.

The heretofore unrecognized existence of clearly cometary images of 313P in SDSS data, despite at least one dedicated effort to search for active comets in that data \citep{sol10}, suggests that renewed efforts to search for MBCs (and comets in general) in SDSS data, taking advantage of recent reprocessing of that data including improved blended source discrimination (Solontoi 2014, private communication), could prove fruitful.  Searches for MBCs in other data archives, especially those obtained by large telescopes or with significant sky coverage, could be similarly productive.

\begin{acknowledgements}
HHH and NH acknowledge support from the NASA Planetary Astronomy program (Grant NNX14AJ38G).
KJM and JK acknowledge support from the NASA Astrobiology Institute under Cooperative Agreement NNA09DA77A issued through the Office of Space Science.
BN is supported by the Ministry of Science of Serbia, Project 176011.
This research utilized Palomar Observatory's Hale Telescope, the Las Cumbres Observatory Global Telescope Network's Faulkes Telescopes, Lowell Observatory's Discovery Channel Telescope, the University of Hawaii 2.2~m telescope, ESO's La Silla Observatory (PID 194.C-0207), and the Canadian Astronomy Data Centre operated by the National Research Council of Canada and the Canadian Space Agency.
SDSS ({\tt http://www.sdss.org/}) is funded by the Alfred P. Sloan Foundation, the Participating Institutions, NSF, the U.S.\ Department of Energy Office of Science, NASA, the Japanese Monbukagakusho, the Max Planck Society, and the Higher Education Funding Council for England, and managed by the Astrophysical Research Consortium for the Participating Institutions.
The Pan-STARRS1 Surveys were made possible by the Institute for Astronomy, the University of Hawaii, the Pan-STARRS Project Office, the Max-Planck Society, the Max Planck Institute for Astronomy, Heidelberg and the Max Planck Institute for Extraterrestrial Physics, Garching, Johns Hopkins University, Durham University, the University of Edinburgh, the Queen's University Belfast, the Harvard-Smithsonian Center for Astrophysics, the Las Cumbres Observatory Global Telescope Network Incorporated, the National Central University of Taiwan, the Space Telescope Science Institute, the Planetary Science Division of the NASA Science Mission Directorate (Grant NNX08AR22G), the National Science Foundation (Grant AST-1238877), and the University of Maryland.  We thank the PS1 Builders and operations staff for PS1's construction and operation,  and access to PS1 data.
\end{acknowledgements}

\end{document}